\documentclass[twocolumn,prb,showpacs]{revtex4}
\usepackage{graphicx}
\usepackage{bm}

\begin{document}
\title{Enhanced magnetocaloric effect in frustrated magnets}
\author{M. E. Zhitomirsky}
\affiliation{European Synchrotron Radiation Facility, BP-220, F-38043
Grenoble, France\\
SPSMS, D\'epartement de Recherche Fondamentale sur la Mati\`ere Condens\'ee,
CEA, F-38054 Grenoble, France\cite{present}}
\date{\today}

\begin{abstract}
The magnetothermodynamics of strongly frustrated classical Heisenberg
antiferromagnets on kagome, garnet, and pyrochlore lattices is examined.
The field induced adiabatic temperature change $(\partial T/\partial H)_S$
is significantly larger for such systems compared to ordinary
non-frustrated magnets and also exceeds the cooling rate of an ideal paramagnet 
in a wide range of fields. An enhancement of the magnetocaloric effect is
related to presence of a macroscopic number of soft modes in frustrated
magnets below the saturation field. Theoretical predictions are confirmed with
extensive Monte Carlo simulations.
\end{abstract}
\pacs{75.30.Sg, 75.10.Hk, 75.50.Ee}
\maketitle

\section{Introduction}

The magnetocaloric effect consists in heating or cooling of matter in
response to variations of external magnetic field. The technique of
adiabatic demagnetization based on the magnetocaloric effect has been
successfully utilized over the years to reach temperatures in a sub-Kelvin
range. \cite{Lounasmaa} Because of its technological simplicity and
independence of gravity the magnetic cooling steadily attracts attention
as a rival to the dilution refrigerators, especially in satellite
applications. \cite{Hagmann,McRae,Kushino} Recent experimental studies of the
magnetocaloric effect in specially designed materials have revealed the 
potential applicability of the magnetic cooling technique for room temperature
refrigeration as well. \cite{Pecharsky,Giguere,Tegus}

Paramagnetic salts, which are standard refrigerant
materials for the low-temperature
magnetic cooling, contain dilute systems of magnetic dipoles. The entropy
of an ideal paramagnet depends on magnetic field $H$ and temperature
$T$ via $H/T$. Therefore, during adiabatic demagnetization,
temperature of a paramagnet decreases {\it linearly} with a field:
$T/ H  = {\rm const}$. The higher the density of the magnetic moments and
their spin number, the greater the cooling power of a refrigerant is. With
increased density of spins, however, the strength of residual interactions
between magnetic dipoles grows and leads to an ordering or a spin glass
transition. The transition temperature $T_c$, which is roughly given by
a strength of magnetic interactions, limits the lowest temperatures achievable
with paramagnetic salts.
Even above $T_c$ the cooling becomes less efficient due to a reduced entropy,
and the adiabatic temperature change is {\it slower} than in an ideal
paramagnet. The cooling rate $(\partial T/\partial H)_S$ of a salt does not,
therefore, exceed the corresponding rate of a paramagnet:
$(\partial T/\partial H)^{\rm para}_S = T/H$.

\begin{figure*}
\includegraphics[width=1.9\columnwidth]{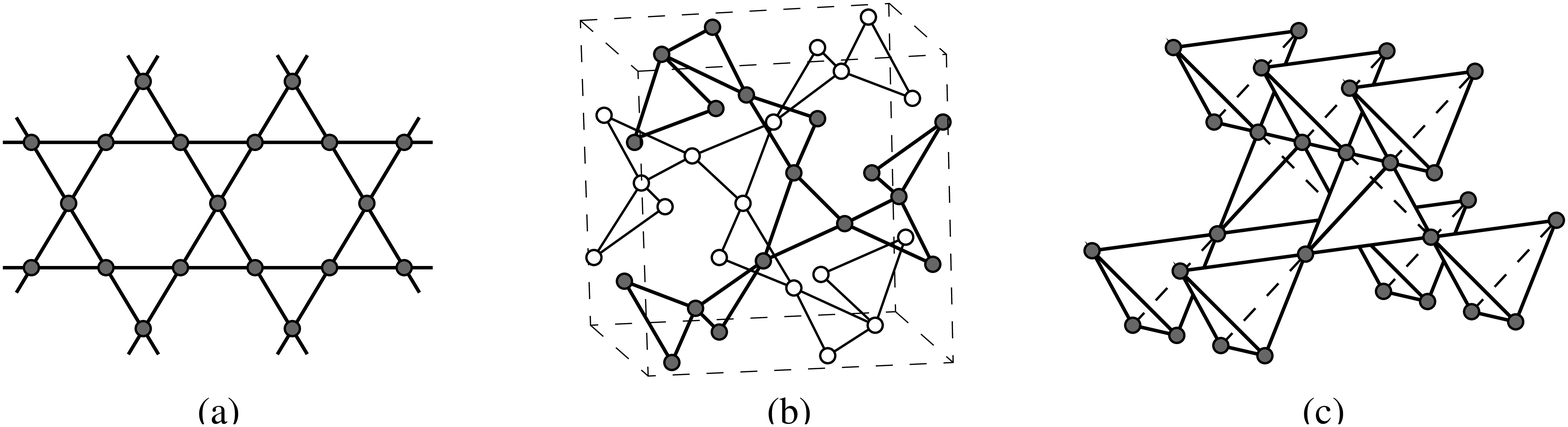}
\caption{\label{Fig1} Magnetic systems with strong frustration: (a)
kagome lattice, (b) garnet lattice, and (c) pyrochlore lattice.}
\end{figure*}

The experimental and theoretical studies over the past decade have
established a new distinct class of magnetic materials called geometrically
frustrated magnets.\cite{Ramirez,Greedan,Moessner} Magnetic ions in
these systems form special types of lattices, Fig.~\ref{Fig1}. Despite
interaction between neighboring spins, strongly frustrated magnets remain
disordered and possess finite entropy at temperatures well below
the Curie-Weiss constant. This property suggests
frustrated magnets as prospective candidates for use in the adiabatic
demagnetization refrigerators. In fact, gadolinium gallium garnet
Gd$_3$Ga$_5$O$_{12}$ (GGG) has long been known as a suitable refrigerant
material. \cite{GGGcool} Recently, GGG with its unique phase diagram in
magnetic field attracted a lot of interest from the point of view of
geometric magnetic frustration on a garnet lattice.\cite{GGGphd} In the
present work we demonstrate a close relation between an enhanced
magnetocaloric effect and strong geometric frustration. In particular,
$(\partial T/\partial H)_S$ in a frustrated magnet can exceed the
cooling rate of an ideal paramagnet by more than an order of magnitude.

Specifically, we study an unusual magnetothermodynamics of classical
Heisenberg antiferromagnets on several frustrated lattices. The choice of
classical models is motivated by large values of spins of the magnetic ions
in frustrated magnets: Cr$^{3+}$ ($S=3/2$), Fe$^{3+}$ ($S=5/2$),
and Gd$^{3+}$ ($S=7/2$).\cite{Ramirez,Greedan} The studied frustrated
structures include two- (2D) and three-dimensional (3D) networks of
corner-sharing triangles known as a kagome, Fig.~\ref{Fig1}a, and a garnet
lattice, Fig.~\ref{Fig1}b, and a 3D network of corner-sharing tetrahedra
of a pyrochlore lattice, Fig.~\ref{Fig1}c. In all three cases the magnetic
Hamiltonian includes exchange interaction of strength $J$ between
the nearest-neighbor classical spins of unit length and the Zeeman energy:
\begin{equation}
\hat{\cal H}=J\sum_{\langle ij\rangle}{\bf S}_i\cdot{\bf S}_j -
{\bf H}\cdot\sum_i{\bf S}_i\ .
\label{Ham}
\end{equation}
The main distinction of strongly frustrated magnets is a large, macroscopic
degeneracy of their classical ground states in zero magnetic field. It arises
due to a so called under-constraint: condition of the minimal energy for
every elementary frustrated unit, triangle or tetrahedron, is not sufficient
to fix all microscopic degrees of freedom. \cite{Ramirez,Moessner}
In applied magnetic field the
macroscopic degeneracy persists up to the saturation field $H_{\rm sat}$,
which is $H_{\rm sat}=6J$ for kagome and garnet antiferromagnets and
$H_{\rm sat}=8J$ for a pyrochlore antiferromagnet. \cite{fafm,kagome}
At zero temperature frustrated magnets loose their degeneracy
and become completely polarized above $H_{\rm sat}$. Transformation 
from the unique ground state at $H>H_{\rm sat}$ to an
infinitely degenerate ground state at $H<H_{\rm sat}$ is accompanied by
condensation of a macroscopic number of zero-energy modes.
\cite{kagome,Honecker} This should be contrasted with the behavior of non- or
weakly-frustrated magnets, where the antiferromagnetic state below
$H_{\rm sat}$ is described by a certain wave-vector and, therefore, the
phase transition at $H=H_{\rm sat}$ corresponds to condensation of one or
a few zero-energy modes. In the vicinity of $H_{\rm sat}$ the thermodynamic
properties of a strongly frustrated magnet are, therefore, controlled
by flat branches of excitations with vanishing energy at $H=H_{\rm sat}$.

The following theoretical analysis is divided into two parts. The next
section gives a general thermodynamic description of magnets with a
macroscopic number of soft excitations. We also calculate the number of
such modes for the three frustrated spin models. The subsequent section
describes Monte Carlo results for various thermodynamic quantities of
frustrated models and also simulations of the adiabatic demagnetization 
process for a few frustrated and non-frustrated magnets.
These are followed by discussion of suitable
magnetic materials, which can exhibit the predicted behavior.

\section{Theory}

Our aim in this section is to understand the effect of soft modes
on a cooling rate of a strongly frustrated magnet under adiabatic
demagnetization. The cooling rate  is related to isothermal
characteristics by
\begin{equation}
\left(\frac{\partial T}{\partial H}\right)_S =
- T \frac{(\partial S/\partial H)_T}{C} \ ,
\label{dTdH}
\end{equation}
where $C$ is the specific heat in a constant magnetic field. We shall
also use the normalized rate  dividing (\ref{dTdH}) by the cooling rate
of an ideal paramagnet $(\partial T/\partial H)^{\rm para}_S=T/H$.
Values $(\partial T/\partial H)^{\rm norm}_S>1$  indicate
that a magnet cools down faster than a paramagnet
in a certain range of fields and temperatures.
The cooling rate is enhanced for systems with large negative
values of $(\partial S/\partial H)_T$.

\subsection{Magnetothermodynamics}

A classical magnet on an $N$-site lattice has $2N$ degrees of freedom
in total with two modes per spin. In the saturated phase above
$H_{\rm sat}$ all $2N$ magnetic excitations are gapped. In the harmonic
approximation the excitation energies of non-frustrated magnets
are given by
\begin{equation}
\omega ({\bf k}) = H - H_{\rm sat} + \varepsilon ({\bf k}) \ ,
\label{quadratic}
\end{equation}
where $\varepsilon ({\bf k})$ is nonnegative and vanishes at one or a few points
in the Brillouin zone. Non-Bravais lattices have several modes for every
wave-vector $\bf k$ and below we imply summation over all such branches.
At low temperatures anharmonic effects are small, which allows to find
explicitly the fluctuation contributions to various
physical quantities. In particular, the specific heat per one spin is
$C=1$ ($k_B\equiv 1$) and
\begin{equation}
\left(\frac{\partial S}{\partial H}\right)_T =
-\frac{1}{2} \sum_{\bf k} \frac{1}{H-H_{\rm sat}+\varepsilon({\bf k})} \ .
\label{dSdH2}
\end{equation}
Since $\varepsilon({\bf k})\sim k^2$ near its zeroes, $(\partial S/\partial H)_T$
at $H=H_{\rm sat}$ is a $T$-independent constant in 3D and has a weak logarithmic singularity
in 2D.

The excitation spectra of strongly frustrated magnets in the saturated phase
are special in that $\varepsilon({\bf k})\equiv 0$ in the whole Brillouin zone for
several flat branches of excitations. We denote the number of such modes as
$N_4$, $N_2$ being the number of ordinary dispersive modes (\ref{quadratic})
and $N_2+N_4=2N$. As field approaches $H_{\rm sat}$, the contribution
from soft dispersionless excitations grows and the harmonic approximation
fails. To study the effect of soft modes one needs to consider an
anharmonic energy functional of the type:
\begin{equation}
E = \sum_{\bf k} (H-H_{\rm sat})|y_{\bf k}|^2 +
\sum_{{\bf k},{\bf p},{\bf q}} 
V_{{\bf k}{\bf p}}^{\bf q} y^*_{\bf k+q}y^*_{\bf p-q}
y_{\bf p} y_{\bf k},
\label{Enonlin}
\end{equation}
where $y_{\bf k}$'s denote collective coordinates of soft modes and
$V_{{\bf k}{\bf p}}^{\bf q}\sim J$ are nonnegative interaction constants.
Cubic terms vanish in the expansion by a spin-rotational symmetry.
Also, an interaction between soft and the harmonic modes can be
neglected at low temperatures. The above energy
functional determines scaling of the corresponding
contribution to the free energy:
\begin{equation}
F_4 = - \frac{N_4}{4}\: T\ln T - T f\left(\frac{H-H_{\rm sat}}{\sqrt{T}}\right).
\label{F4}
\end{equation}
The function $f(u)$ has the following asymptotes:
$f(u)\approx f(0) + u f'(0)$ with $f(0)>0$ and $f'(0)<0$ as $u\rightarrow 0$,
and $f(u) \approx -(N_4/2)\ln u$ as $u\rightarrow \infty$.

It follows from Eq.~(\ref{F4}) that 
the specific heat and the magnetic susceptibility depend
on magnetic field and temperature 
only via combination $(H-H_{\rm sat})/\sqrt{T}$.\cite{kagome}
In particular, the specific heat per one spin reaches a universal
fractional value at $H=H_{\rm sat}$: $C=(\frac{1}{2}N_2+\frac{1}{4}N_4)/N
= 1 - N_4/(4N)$. For the magnetocaloric effect, Eq.~(\ref{F4}) gives
at $H=H_{\rm sat}$:
\begin{equation}
\left(\frac{\partial S}{\partial H}\right)_T \propto -
\frac{N_4}{\sqrt{T}} \ ,
\label{dSdH4}
\end{equation}
which significantly exceeds the contribution of dispersive modes
Eq.~(\ref{dSdH2}). Thus, condensation of a macroscopic number of zero-energy
modes at the saturation field greatly enhances the magnetocaloric effect in
strongly frustrated magnets at low temperatures. An important characteristics
of a magnet is, then, the number of soft modes $N_4$.

\subsection{Soft modes}

\subsubsection{Kagome antiferromagnet}

In order to find the number of soft modes we perform classical spin-wave
calculations in the saturated collinear state. A primitive unit cell of a
kagome lattice contains one triangular plaquette. The corresponding Bravais
lattice has $N/3$ states. Expanding spin components along the field in 
small transverse deviations we obtain a classical spin-wave Hamiltonian to
a desired order. In the harmonic approximation two transverse polarizations
for the spin excitations are independent. As a result, the spectrum consists
of three double-degenerate modes:
\begin{eqnarray}
\omega_{1}({\bf k}) & = & H - 6J \ ,  \nonumber \\
\omega_{2,3}({\bf k})& = & H - 3J \pm J\sqrt{3(1+2\gamma_{\bf k})} \ ,
\label{freq_kag}
\end{eqnarray}
where $\gamma_{\bf k} = \frac{1}{3}(\cos k_x + 2\cos\frac{1}{2}k_x
\cos\frac{\sqrt{3}}{2}k_y)$. The total number of soft quartic modes at
$H_{\rm sat}=6J$ is, therefore, $N^{\rm kag}_4=2N/3$. This is exactly
twice the number of hexagon voids on a kagome lattice. The
correspondence between such voids and soft modes has first been found
for a kagome antiferromagnet in zero field \cite{Chalker} and was,
later, extended to finite magnetic fields.\cite{kagome} Local soft modes
in the collinear state at $H=H_{\rm sat}$ consist in alternate tilting of
spins around elementary hexagons in two different directions perpendicular
to the field. The geometric picture of local soft modes can be
straightforwardly generalized for a classical spin model on a
checker-board lattice, which is the other planar frustrated lattice
with voids around empty squares.\cite{Honecker}
The checker-board antiferromagnet has $N$ local zero-energy modes at
$H_{\rm sat}=8J$.

\begin{figure}[b]
\includegraphics[width=0.6\columnwidth]{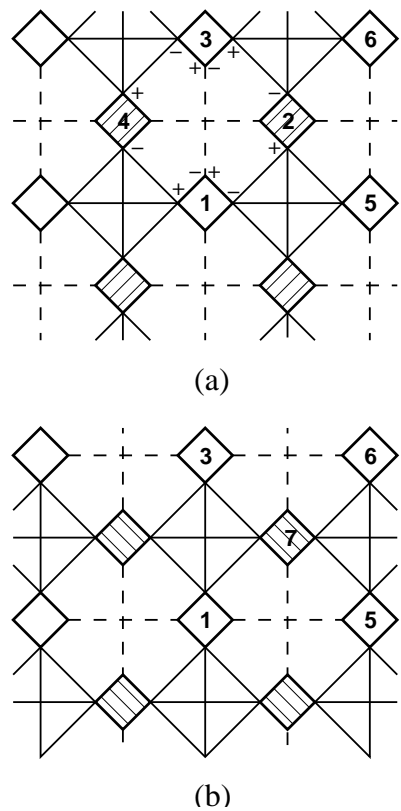}
\caption{\label{Fig2} Two projections of a pyrochlore lattice on a (001)
plane. Small, light and dashed, squares represent one type of tetrahedra;
big squares with crosses correspond to the other type of tetrahedra. 
Small light
tetrahedra belong to the same $(001)$ plane of an fcc lattice. Small dashed
tetrahedra lie in an adjacent plane: (a) below and (b) above.
Soft spin modes on hexagon loops $l_{123}$ and $l_{134}$ are shown
with $+$'s and $-$'s on the top panel.}
\end{figure}

\subsubsection{Pyrochlore antiferromagnet}

The spin-wave calculations for a pyrochlore antiferromagnet
yield four double-degenerate modes defined in the Brillouin
zone of a face-centered cubic (fcc) lattice, which has $N/4$
states:
\begin{eqnarray}
\omega_{1,2}({\bf k}) & = & H - 8J \ , \nonumber \\
\omega_{3,4}({\bf k}) & = & H - 4J \pm 2J\sqrt{1+3\nu_{\bf k}}
\label{freq_pyro}
\end{eqnarray}
with $\nu_{\bf k} = \frac{1}{6}(\cos k_x + \cos k_y +
4\cos k_z \cos \frac{1}{2}k_x\cos\frac{1}{2}k_y)$. The total
number of zero modes in a pyrochlore antiferromagnet at the saturation
field is $N^{\rm pyr}_4=N$, {\it i.e.}\ the same as for the checker-board
model. The geometric interpretation of these modes becomes, however,
more complicated. A dispersionless soft mode can still be attributed to
a local perturbation in the collinear structure:
alternate tilting of spins around a closed line,
which shares two vertices with each crossed tetrahedron.
The shortest such loop on a pyrochlore lattice is a hexagon.
Simple counting yields $N$ hexagons for an $N$-site pyrochlore lattice.
If a one-to-one correspondence
between the soft modes and the hexagons holds, then the
above number would imply $2N$ soft modes, which is twice more
than the spin-wave prediction. The contradiction is
resolved in favor of the spin-wave result by establishing that
some of the hexagon loops on a pyrochlore lattice are linearly
dependent with total $N/2$ linear relations. This brings
the number of independent hexagon spin-loops (modes) to $N/2$
in agreement with the spin-wave expansion. Below, we briefly
sketch a proof of the above statement.

A pyrochlore lattice, Fig.~\ref{Fig1}c, is formed with two types of tetrahedra.
To simplify geometric representation we draw a pyrochlore structure
as an fcc lattice constructed from one type of tetrahedra. A cut through such
a lattice along $(001)$ plane is shown in Fig.~\ref{Fig2}. Hexagons in
Fig.~\ref{Fig2} correspond to closed loops which include two tetrahedra in
one layer and one tetrahedron in an adjacent layer. 
Soft spin modes on two such loops are shown in Fig.~\ref{Fig2}
with pluses and minuses, which indicate alternate tilting
of spins from the field direction.
The first type of a linear
relation between spin-loops appears for a block formed by four
tetrahedra numbered 1--4 in Fig.~\ref{Fig2}a. A superposition of 
the local soft modes corresponding to hexagon
loops $l_{123}$ and $l_{134}$ leaves two inner spins unmoved and tilts
only outer spins around the block perimeter. The same 
eight-spin mode is obtained by a
superposition of spin-loops $l_{124}$ and $l_{234}$. Thus, every such block of
four tetrahedra provides one linear relation between four loops.
This reduces the number of independent spin-loops by the number of such 
blocks $N/4$. The second type of a linear relation corresponds to
a block of six tetrahedra: four in one layer, 1, 3, 5, and 6, and two
in adjacent layers, 2 and 7, Figs.~\ref{Fig2}a,b.
A superposition of spin-loops $l_{123}$ and $l_{256}$ corresponds to
the same mode as a superposition of $l_{157}$ and $l_{367}$.
Counting all such blocks brings additional $N/4$ linear relations.
This makes in total $N/2$ linear relations between
hexagonal spin-loops and, hence, $N^{\rm pyr}_4 = 2(N-N/2)=N$ soft modes for
a pyrochlore antiferromagnet at the saturation field.
Composite low-energy excitations on six-spin loops, which are similar
to the above soft modes, have been recently observed
in neutron scattering studies of pyrochlore antiferromagnet
ZnCr$_2$O$_4$ in zero field.\cite{Broholm}

\subsubsection{Garnet antiferromagnet}

The structure of a garnet lattice is by far the most complicated among
frustrated lattices. In GGG the magnetic Gd$^{3+}$ ions form two
interpenetrating triangular sublattices with 24 atoms per a cubic unit cell.
Exchange interaction couples each spin with its nearest-neighbors from the
same sublattice. Spins on two different sublattices interact only via weaker
dipolar forces. In the present study we focus on the behavior of an exchange
model and, therefore, consider only one sublattice of corner-sharing
triangles with 12 spins per a unit cell represented, e.g., by dark circles
in Fig.~\ref{Fig1}b. The nearest-neighbor exchange Hamiltonian on a garnet
lattice can be written as
\begin{widetext}
\begin{eqnarray}
{\cal H}  & = &J \sum_i \Bigl[
{\bf S}_{1i}\cdot({\bf S}_{2i}+{\bf S}_{3i}+{\bf S}_{4i+x}+{\bf S}_{5i+x}) +
{\bf S}_{2i}\cdot({\bf S}_{3i}+{\bf S}_{6i+y}+{\bf S}_{7i+y}) +
{\bf S}_{3i}\cdot({\bf S}_{8i+z}+{\bf S}_{9i+z})
\nonumber\\
& & \mbox{} +
{\bf S}_{4i}\cdot({\bf S}_{5i}+{\bf S}_{8i+z}+{\bf S}_{10i+z}) +
{\bf S}_{5i}\cdot({\bf S}_{6i-x}+{\bf S}_{11i}) +
{\bf S}_{6i}\cdot({\bf S}_{7i}+{\bf S}_{11i+x}) +
{\bf S}_{7i}\cdot({\bf S}_{9i-y}+{\bf S}_{12i}) \nonumber\\
& & \mbox{} +
{\bf S}_{8i}\cdot({\bf S}_{9i}+{\bf S}_{10i}) +
{\bf S}_{9i}\cdot {\bf S}_{12i+y} +
{\bf S}_{10i}\cdot({\bf S}_{11i}+{\bf S}_{12i}) +
{\bf S}_{11i}\cdot {\bf S}_{12i}  \Bigr] + {\cal H}_{\rm Zeeman} \ ,
\end{eqnarray}
\end{widetext}
where the first spin index $j=1$--12 is the site number in a unit
cell and the second spin index $i$ is the cell number on a cubic
lattice; $i\pm x$,... being  adjacent cells along three
orthogonal directions.

After Fourier transformation the spin-wave energies are given by eigenvalues
of a $12\times 12$ matrix. The corresponding equation factorizes into
a product of $(\omega-H+6J)^4$ and an eighth-order polynomial in
$\omega$. \cite{math} At $H=6J$ there are exactly four zero roots for an
arbitrary value of the wave-vector. Zero eigenvalues determine
$N^{\rm gar}_4=2N/3$ soft modes at the saturation field, which is the same
number as for a kagome antiferromagnet. Thus, the term {\it hyper}-kagome used
sometimes to refer to a garnet structure receives an extra justification.
Geometric picture of soft modes in a hyper-kagome
antiferromagnet is, however, much more complicated than in its planar
counterpart. The shortest closed loop on a garnet lattice, which corresponds
to an elementary soft mode, crosses ten triangles. At the same time each bond
between a pair of nearest-neighbor spins participates in five such loops.
\cite{private} This should be contrasted with simple hexagon loops on a kagome
lattice, where every bond is included in one loop only.  The total number of
ten-triangle loops for a garnet lattice is calculated to be $N$. A comparison
with $N/3$ soft modes for one transverse polarization found in the spin-wave
analysis suggests that there are $2N/3$ linear relations between closed loops
for a garnet structure. An underlying geometric picture behind these relations
remains to be clarified.

To conclude this section we note that a pyrochlore lattice is the most
frustrated one among the above three structures, because it has the largest
number of soft modes. The same is also true in zero magnetic field, where
a classical Heisenberg pyrochlore lattice 
antiferromagnet remains disordered down to
lowest temperatures, \cite{Moessner_Chalker} whereas a classical kagome
antiferromagnet develops a unique planar triatic order parameter.
\cite{kagome,Chalker} As a result, a pyrochlore antiferromagnet should
exhibit the fastest cooling rate under adiabatic demagnetization among
frustrated magnets.

\begin{figure}[t]
\includegraphics[width=0.8\columnwidth]{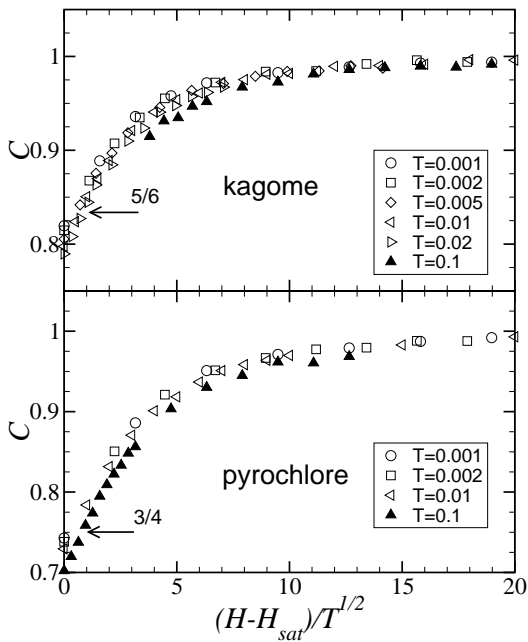}
\caption{\label{Fig3}
Scaling behavior of the specific heat in the saturated phase for a kagome
(top panel) and a pyrochlore (bottom panel) antiferromagnet. Different
symbols correspond to field-scans performed at several fixed temperatures.
Arrows indicate the limiting values of the specific heat per one site
at the saturation field. Fields and temperatures are given in units of $J$.}
\end{figure}

\section{Monte Carlo simulations}

Monte Carlo (MC) simulations of the three spin models have been performed
with the standard Metropolis algorithm. To improve statistics at low
temperatures we introduce a maximum change $\Delta S^z=3T$ for a choice
of a new spin direction in the local coordinate frame. In this way an acceptance
rate for spin moves is kept at the level of 30\%. (Other values
$\Delta S^z/T\sim 1$--5 will work as well.) MC simulations are done
on finite clusters with periodic boundary conditions. Typical cluster
sizes are $N=3\times18^2=972$ for a kagome lattice, 
$N=4\times6^3=864$ for a pyrochlore lattice, and 
$N=12\times5^3=1500$ for a garnet lattice. Comparison with smaller and larger
clusters have been made to insure that finite-size corrections are small.
Typical MC runs are carried with $10^5$ MC steps per spin for equilibration
followed by averaging over $10^5$ measurements made in intervals of 5--10 MC
steps. Statistical errors are estimated by performing several ($\sim 5$)
runs from different initial configurations. The errors for all presented
data are small and do not exceed symbol sizes. The specific heat is calculated
from the variance of the total energy: $C=(\langle E^2\rangle -
\langle E\rangle^2)/T^2$, while the magnetocaloric effect is expressed via
a cumulant of the energy and the magnetization: $(\partial
S/\partial H)_T = (\langle EM\rangle-\langle E\rangle\langle M\rangle)/T$.

\begin{figure}
\includegraphics[width=0.8\columnwidth]{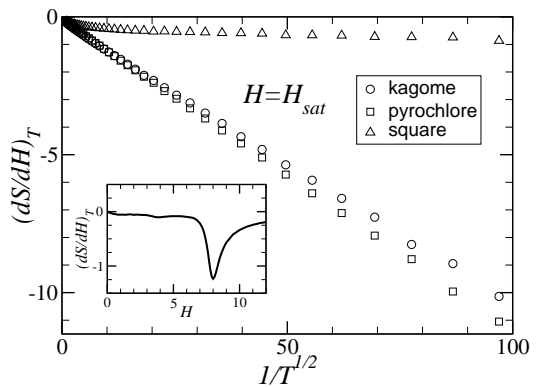}
\caption{\label{Fig4}
Scaling behavior of $(\partial S/\partial H)_T$ at $H=H_{\rm sat}$
in kagome and pyrochlore antiferromagnets. Inset shows the field
dependence of $(\partial S/\partial H)_T$ at $T=0.01J$ for a pyrochlore
antiferromagnet. Fields and temperatures are given in units of $J$.}
\end{figure}

We present in Fig.~\ref{Fig3} the specific heat per one site of kagome and
pyrochlore antiferromagnets plotted with respect to the scaling parameter
$u = (H-H_{\rm sat})/\sqrt{T}>0$. The data have been collected during
field-scans at fixed temperatures. Points from different scans, except of
the highest temperature of $T=0.1J$, fall on the same curve confirming
the predicted scaling form. The specific heat at $H=H_{\rm sat}$ ($u=0$)
reaches values which are close to the asymptotic results derived from
the number of soft quartic modes: $C=5/6$  and $C=3/4$ for a kagome and
a pyrochlore antiferromagnet, respectively. The magnetocaloric effect
near saturation is presented in Fig.~\ref{Fig4}. The main plot shows
temperature dependence of $(\partial S/\partial H)_T$ exactly at the
saturation field for kagome, pyrochlore and square lattice
(576-site cluster) antiferromagnets.
Numerical results for a garnet antiferromagnet are not included
because at $H>H_{\rm sat}$ they virtually coincide with the data for
a kagome antiferromagnet. Significant difference (Fig.~\ref{Fig4})
between the frustrated magnets on one side and a non-frustrated
square lattice antiferromagnet on the other side confirms an enhancement
of the magnetocaloric effect related to magnetic frustration.
A range of magnetic fields in the vicinity of $H_{\rm sat}$,
where such an enhancement occurs, is illustrated on the inset.

\begin{figure}
\includegraphics[width=0.8\columnwidth]{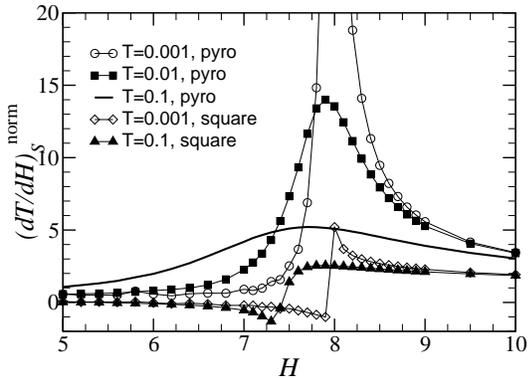}
\caption{\label{Fig5}
Cooling rates of pyrochlore and square lattice
antiferromagnets normalized
to a cooling rate of an ideal paramagnet.
Fields and temperatures are given in units of $J$.}
\end{figure}

Direct comparison of the cooling rates of magnets with and without
frustration is made in Fig.~\ref{Fig5}, where results for a pyrochlore
and a square lattice antiferromagnet are plotted for several field scans.
The two magnets have the same saturation field $H_{\rm sat}=8J$.
For a pyrochlore antiferromagnet this field corresponds to a
crossover from the nondegenerate saturated phase into a highly
degenerate state below $H_{\rm sat}$. The crossover is accompanied
by a large increase in the adiabatic temperature variation, which exceeds
the cooling rate of an ideal paramagnet by more than an order of magnitude.
In contrast the spin model on a square lattice has a real transition near
$H_{\rm sat}$ into a nondegenerate antiferromagnetic state with divergent
correlation length. The cooling rate becomes negative below
the transition, which means that a magnet heats up during demagnetization
rather than cooling down. There is also no significant enhancement of
the cooling rate above $H_{\rm sat}$ apart from a small narrow peak due to
an anomaly in $(\partial S/\partial H)_T$ at the transition.

\begin{figure}[t]
\includegraphics[width=0.8\columnwidth]{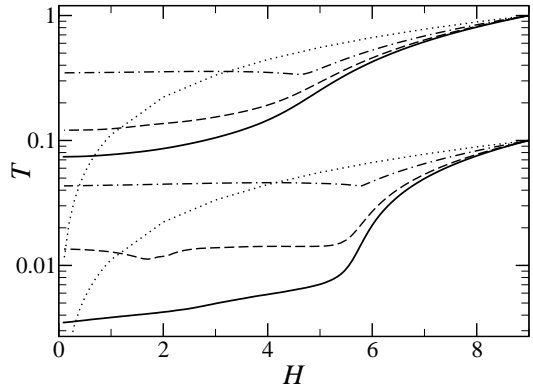}
\caption{\label{Fig6}
Temperature variation during adiabatic demagnetization
of a pyrochlore (full line), a garnet (dashed line),
and a cubic (dot-dashed line) antiferromagnet. Dotted lines represent
demagnetization of an ideal paramagnet $(T/H={\rm const})$.
Fields and temperatures are given in units of $J$.}
\end{figure}

Finally, a full adiabatic demagnetization process can be simulated
with MC by measuring at fixed $T$ and $H$ isothermal characteristics
$C$ and $(\partial S/\partial H)_T$ to find the cooling rate (\ref{dTdH})
and, then, choosing a new value of temperature according to
$\Delta T_{ad}/\Delta H=(\partial T/\partial H)_S$ for sufficiently small
$\Delta H$. Adiabatic temperature changes for demagnetization from two
starting temperatures $T=J$ and $T=0.1J$ and the same initial field
$H=9J$ are shown for garnet, pyrochlore, and cubic ($N=512$) antiferromagnets
in Fig.~\ref{Fig6}. Since saturation fields for the three magnets are
different, we use rescaled exchange constants $J^*=0.75J$ for a pyrochlore and
$J^*=0.5J$ for a cubic antiferromagnet to set the common saturation field of
$H_{\rm sat}=6J$ in all three cases. During demagnetization the non-frustrated
cubic antiferromagnet has the smallest relative temperature variation
and starts to heat up below the transition into antiferromagnetically
ordered state. The adiabatic temperature decrease for frustrated magnets is
much larger with the largest effect exhibited by a pyrochlore antiferromagnet.
Note, that temperature of a pyrochlore antiferromagnet drops by more than
ten times as field decreases from $H_{\rm ini}=1.5H_{\rm sat}$ to
$H_{\rm fin}=0.8H_{\rm sat}$. To achieve a similar cooling with a paramagnetic
salt magnetic field has to be decreased at least by a factor of ten.
The two distinctive features of the demagnetization of a pyrochlore
antiferromagnet are (i) a persistent temperature decrease up to a zero field,
though with a reduced cooling rate below $H_{\rm sat}$, and (ii) an increase
in the relative temperature variation as a starting temperature goes down,
which is a consequence of the low-temperature singularity  (\ref{dSdH4}).

\section{Discussion}

The practical limitation on experimental observation
of an enhanced magnetocaloric effect comes from a strength
of the saturation field.
At present, only frustrated magnets with
exchange constants of the order of few Kelvin or less
can be saturated with dc magnetic fields. The prime example
is gadolinium gallium garnet with $H_{\rm sat}\simeq 1.7$~T,\cite{GGGphd}
which is a popular refrigerant material  in an applied experimental
research on the magnetic cooling. \cite{GGGcool,Kushino}
Quantitative comparison of the cooling rates of
an ideal paramagnet
and a Heisenberg garnet antiferromagnet 
is made in Fig.~\ref{Fig6}. Starting with the same
temperature a garnet antiferromagnet cools down to much lower 
temperatures than a paramagnet in a wide range of magnetic fields.
There are also two pyrochlore antiferromagnets Gd$_2$Ti$_2$O$_7$ and 
Gd$_2$Sn$_2$O$_7$,\cite{Raju,Ramirez02,Bramwell} which have 
saturation fields
about 7~T. It is an interesting experimental challenge to observe
the fast temperature decrease upon adiabatic demagnetization in these
materials. Operation with relatively high magnetic fields makes rather
questionable a technological application of the two gadolinium pyrochlores.
The theoretical results suggest, though, a new direction for a search
of suitable materials: a pyrochlore antiferromagnet
with a factor of four smaller exchange constant than
in Gd$_2$Ti$_2$O$_7$ would be a more efficient refrigerant than
Gd$_3$Ga$_5$O$_{12}$, while operating in a similar
range of fields and temperatures. 

Real materials have, of course, more complicated magnetic
interactions than just a nearest-neighbor exchange
between classical spins. For example, in GGG a dipole-dipole
interaction between neighboring spins amounts 
to 20--30\% of the corresponding exchange energy.\cite{GGGphd}
Long-range dipolar interactions play also a significant
role in the two pyrochlore compounds.\cite{Raju,Ramirez02}
Influence of these and other extra perturbations on
the magnetocaloric effect in frustrated magnets requires further
theoretical analysis.  Meantime, one can make several
general remarks on the relative importance of various
perturbations. The quantum effects omitted in the present
study do not destroy the anomalous magnetothermodynamics 
near $H_{\rm sat}$. Similar to their classical counterparts 
quantum antiferromagnets on the three
frustrated lattices also have flat branches of excitations.
\cite{Honecker} 
The anomaly (\ref{dSdH4}) is, nevertheless, affected
by a weak dispersion ($\sim J_{\rm disp}$) of soft
excitations induced, e.g., by a second-neighbor exchange
or dipole-dipole interactions. The magnetocaloric effect
will saturate below $T^*\sim J_{\rm disp}$ at 
$(\partial S/\partial H)_T \sim 1/\sqrt{T^*}$, which could be
relatively large. Such a frustrated magnet can still be used, though less
efficiently, as a refrigerant to reach temperatures below $T^*$.
A further analysis of a destructive role of various
perturbations should focus, therefore, on an induced dispersion
in flat magnon branches.

\begin{acknowledgments}
I am greatly indebted to T. M. Rice, who's questions have inspired this study.
In the course of the work I benefited from fruitful discussions with
A. I. Golov, C. L. Henley, A. Honecker, A. R. Muratov, and O. A. Petrenko.
\end{acknowledgments}


\begin{thebibliography}{99}
\bibitem{math}
we use the {\sc Mathematica} software for this calculation.
\end{thebibliography}
\end{document}